\def\gsim{~\rlap{$>$}{\lower 1.0ex\hbox{$\sim$}}}
\def\lsim{~\rlap{$<$}{\lower 1.0ex\hbox{$\sim$}}}

\def\h2o{\rm{H_{2}O}}
\def\mh2{\rm{H_{2}}}

\def\co2{\rm{CO_{2}}}
\def\ch4{\rm{CH_{4}}}
\documentclass[12pt, preprint]{aastex}
\usepackage{graphicx}
\usepackage{natbib,color,lscape}
\usepackage{subfigure}
\usepackage{threeparttable}
\usepackage{url}
\usepackage{subfigure}
\usepackage{threeparttable}
\usepackage{rotating}

\begin{document}
\title{HABITABLE ZONES AROUND MAIN-SEQUENCE STARS: DEPENDENCE ON PLANETARY MASS}
\author{Ravi kumar Kopparapu\altaffilmark{1,2,3,4,5}, 
        Ramses M. Ramirez\altaffilmark{1,2,3,4}, 
        James SchottelKotte\altaffilmark{6},
        James F. Kasting\altaffilmark{1,2,3,4},
        Shawn Domagal-Goldman\altaffilmark{2,7},
        Vincent Eymet\altaffilmark{8}}
\altaffiltext{1}{Department of Geosciences, Penn State University, 443 
Deike Building, University Park, PA 16802, USA}
\altaffiltext{2}{NASA Astrobiology Institute's Virtual Planetary Laboratory, P.O. Box 351580, Seattle, WA 98195, USA}
\altaffiltext{3}{Penn State Astrobiology Research Center, 2217 Earth and Engineering Sciences Building
University Park, PA 16802}
\altaffiltext{4}{Center for Exoplanets \& Habitable Worlds, The Pennsylvania State University, University
Park, PA 16802}
\altaffiltext{5}{Blue Marble Space Institute of Science, PO Box 85561, Seattle, Washington 98145-1561, USA}
\altaffiltext{6}{Department of Astronomy \& Astophysics, The Pennsylvania State University, 525 Davey
Laboratory, University Park, 16802, USA}
\altaffiltext{7}{Planetary Environments Laboratory, NASA Goddard Space Flight Center}
\altaffiltext{8}{Laboratoire d'Astrophysique de Bordeaux, Universite de Bordeaux 1, UMR 5804}

\begin{abstract}
The ongoing discoveries of extrasolar planets are unveiling  a wide range of terrestrial mass (size)
planets around their host stars. 
 In this letter, we present estimates of habitable zones (HZs) around stars with 
stellar effective temperatures in the range $2600$ K - $7200$ K, 
for planetary masses between $0.1$ M$_{\oplus}$ and $5$ M$_{\oplus}$.  Assuming $\h2o$- (inner HZ) and 
$\co2$- (outer HZ) dominated atmospheres,  and scaling the  
background N$_{2}$ atmospheric pressure with the radius of the planet,  
our results indicate that larger planets have 
wider HZs than do smaller  ones.   Specifically, with the assumption that 
smaller planets will have less dense atmospheres, the inner edge of the HZ 
(“runaway greenhouse” limit) moves outward ($\sim 10 \%$ lower than Earth flux) for low mass planets due to 
larger greenhouse effect arising from  the increased $\h2o$ column depth. 
 For larger planets, the $\h2o$ column depth is smaller, and higher
temperatures are needed before water vapor completely dominates the outgoing longwave radiation.
Hence the inner edge moves inward ($\sim 7 \%$ higher than Earth's flux).
The outer HZ changes little due to the competing effects of the greenhouse effect and an increase
in albedo.
 New, 3-D climate model results from other groups are also summarized, and we argue that further, independent studies are needed to verify
their predictions.
Combined with our previous work, 
the results presented here provide refined estimates of HZs 
around main-sequence stars and provide a step towards a more comprehensive analysis of HZs.

\end{abstract}

\maketitle

\section{Introduction}
\label{intro}
Recent observational surveys have discovered several potential habitable zone (HZ) planet candidates
\citep{Udry2007, Vogt2010, Pepe2011a, Borucki2011, Bonfils2011, Borucki2012, Vogt2012, Tuomi2012b,
Anglada-Escude2013}, and it is
expected that this number will greatly increase as time passes \citep{DC2013, KoppM2013,Gaidos2013}.
 Accordingly, the circumstellar HZ is defined as the  region around which a
terrestrial mass planet, with favorable
atmospheric conditions, can sustain liquid water on its surface
\citep{Huang1959, Hart1978, Kasting1993, Selsis2007b,Kopp2013}.
Currently, more than $1600$ extra-solar planetary systems have been
detected and  $> 2700$ additional candidate systems
from the {\it Kepler} mission are waiting to be confirmed \citep{Batalha2013,Lissauer2014, Rowe2014}.

%Most of the HZ limits that were cited in recent  discoveries were based on
% 1-D radiative-convective, cloud-free climate model 
%calculations by \cite{Kasting1993}. %Several other studies \citep{Underwood2003, Selsis2007b} 
%parameterized these results to estimate relationships between HZ boundaries and stellar parameters 
%for stars of different spectral types.
Recently \cite{Kopp2013} obtained new, improved estimates of the boundaries of the HZ by 
updating \cite{Kasting1993} model with new $\h2o$ and $\co2$ absorption coefficients 
from updated line-
by-line (LBL) databases such as HITRAN 2008  \citep{Rothman2009} and HITEMP 2010 \citep{Rothman2010}.
%According to their revised model,  a conservative estimate of the inner HZ (IHZ) for our Sun is at
%$0.99$ AU and the outer HZ (OHZ) is at $1.70$ AU. These values represent the 
%``water loss'' (moist greenhouse) and ``maximum greenhouse'' limits, respectively. 

Several other recent studies used 3D global circulation models (GCMs) to study the potential habitability of specific systems
\citep{Wordsworth2010,Forget2013}. Specifically, a recent study by \cite{Yang2013}
proposed that stabilizing cloud feedback can expand
the inner HZ (IHZ) to roughly twice the stellar flux found from 1D climate calculations 
for tidally locked planets or planets 
that are in synchronous rotation around 
low mass stars. The stabilizing feedback arises from an increase in the planetary albedo due to the
presence of
thick water clouds at the sub-stellar point.  In contrast,  \cite{Leconte2013} found that for a rapidly rotating planet
similar to Earth around a Sun-like star, clouds have a {\it destabilizing} feedback on the 
long-term warming. This is because of the displacement of the cloud formation layer to higher altitudes, 
increasing the greenhouse effect
of the clouds compared to the cooling effect caused by their albedo.
 While clouds provide a positive feedback in their model, \cite{Leconte2013} show
that Earth's troposphere is not saturated everywhere,
 and that these unsaturated regions radiate efficiently to space, thereby cooling the planet. 
Consequently, they find that the IHZ is
closer to the Sun, at $0.95$ AU, 
than predicted by the 1-D model of
\cite{Kopp2013}. A similar study by \cite{WT2013}, using the 3D Community Atmosphere Model 3 (CAM3),
 also found that the inner edge can be as close as $0.93$ AU for our Sun. 
These results highlight the importance of 3D GCMs in understanding the varying climate feedacks associated with both
tidally locked and rapidly rotating planets. Further studies using 3D models will be necessary to obtain
a consensus on the location of the inner edge of the HZ.

%Although both the 1D and 3D models estimated the HZ boundaries,
%the focus of previous work was restricted to Earth-mass planets. Current discoveries of 
%extrasolar terrestrial planets span a wide range of mass and radius,
%and it is important to consider the effect of varying planetary mass/size on the estimates of the HZs.

        Here, we 
	consider planetary masses $M_{p}$ between  $0.1$ M$_{\oplus}$ $\le M_{p} \le 5$ M$_{\oplus}$.
 The lower limit includes Mars-mass planets. The upper limit is based on the observation that the theoretical and observed 
mass-radius relationships have different slopes beyond $5$M$_{\oplus}$(see \S\ref{sec2}), 
suggesting the accumulation of an increasingly significant gas envelope for planets with sizes larger 
than $5$M$_{\oplus}$. 
%As in \cite{Kopp2013}, our model assumes that the 
%carbonate-silicate weathering feedback (which requires continents) is active. Using this approach, we derive the HZ limits 
%around main-sequence stars with effective temperatures ($T_{eff}$) between $2600$ K and $7200$ K.

%	the higher limit is based on the observation that the theoretical and observed mass-radius relationships 
%have different slopes beyond $5$M$_{\oplus}$(see \S\ref{sec2}). Recent studies indicated that a tectonically active
%%super-Earth planet can have exposed continents, which can enable silicate weathering process \citep{CA2014}.
%% a theoretical limit beyond which gas-giant planet formation is 
%%thought to occur.
%As in \cite{Kopp2013}, our model assumes that the carbonate-silicate weathering process
%is active, and we derive the HZ limits  around main-sequence stars with 
%	 effective temperatures ($T_{eff}$) between $2600$ K and $7200$ K. 
%A parametric fit relating
%	the HZ fluxes with planetary mass and $T_{eff}$ is provided in the form of a FORTRAN code, which
%	can be downloaded from the online version of the Journal.

	 The outline of the paper is as follows: In \S\ref{sec2} we briefly describe
	our 1-D cloud-free climate model.
	In \S\ref{results} we present results from our climate model and illustrate various HZ limits as 
	a function of planetary mass.   In \S\ref{discuss}, we
provide an analytical equation to calculate HZs incorporating various 3D GCM results. We conclude in
	\S\ref{conclusions}.

	\section{Model Description}
	\label{sec2}
	We used a 1D, radiative-convective, cloud-free climate model from
	\cite{Kopp2013}.
%, which was developed as an update to the \cite{Kasting1993} model. 
%Major improvements include
%	new $\h2o$ and $\co2$ absorption coefficients derived from the HITRAN 2008 and HITEMP
%	2010 line-by-line databases.
	 We considered  planets of masses  $0.1$ M$_\oplus$ and
	 $5$ M$_\oplus$, which were assumed to have $\h2o$-(IHZ) or $\co2$-
	(OHZ) dominated atmospheres  with N$_{2}$ as a background gas. 
 We explored the following cases: (1) N$_{2}$ partial pressure ($pN_{2}$) was varied for a fixed planet mass 
($1$ M$_\oplus$) to study the effect of non-condensable background gas on the HZ limits, (2) 
N$_{2}$ background pressure was fixed at a low value of $0.01$ bar for various planetary masses 
($0.1, 1$ and $5$ M$_\oplus$) to study the effect of gravity alone and (3) N$_{2}$ 
pressure was scaled according to the planetary radius, accounting implicitly for the possible 
effect of planet size on volatile abundance.

        For the last case, we assume that the amount of volatiles acquired by a planet during 
the late stages of its formation is 
proportional to the planet's mass. We further assume that the fraction of these volatiles that are 
outgassed either during or after accretion is the same for all planets. 
 We should caution that volatile delivery to a planet is stochastic in nature, and may be a weak 
function of planetary mass
\citep{Raymond2006,Raymond2007}. 
%Variations in the volatile content can be stochastic due to the 
%amount of accreting embryos, initial location of the planet and the influence of any giant planet(s) in the
%system.  
Still, this is the best assumption we can make in the absence of a 
rigorous theory of how planetary volatile content varies with planet mass.

The surface pressure, $P_{s}$, of 
a planet for this last case is then given by:
\begin{eqnarray}
\frac{P_{s}}{P_{s}^{o}} &=& \frac{N_{col}}{N_{col}^{o}}~.~\frac{g}{g_{o}}
\label{press}
\end{eqnarray}
where $N_{col}$ is the N$_{2}$ atmospheric column mass density, which is atmospheric mass 
(scales with planetary mass, $M_{p}$) 
divided by the surface area of the planet\footnote{In general, $N_{col} = \int_{z1}^{z2} \rho(z) ~.~ dz$, where 
$z$ is the atmospheric height, $\rho(z)$ is the mass density of the atmosphere. In essence, $N_{col}$ is the 
mass per unit area of a column of atmosphere.}, 
and $g$ is the acceleration due to gravity. $P_{s}^{o}$, $N_{col}^{o}$ and $g_{o}$ are the corresponding
values for Earth.

Both the terms on the right-hand side of Eq.(\ref{press}) are proportional to $M_{p}/R_{p}^{2}$, where $R_{p}$ 
is the radius of the planet. Therefore, Eq.(\ref{press}) can be written as:
\begin{eqnarray}
\frac{P_{s}}{P_{s}^{o}} &=& \biggl(\frac{M_{p}}{M_{o}}\biggr)^{2}~.~\biggl(\frac{R_{o}}{R_{p}}\biggr)^{4}
\label{press2}
\end{eqnarray}

%Because planet mass, $M_{p}$, 
%scales approximately as radius cubed, whereas surface area scales as radius squared, the N$_{2}$
%atmospheric column density (or column mass)
%%, which is atmospheric mass (scales with planetary mass) 
%%divided by the surface area of the planet,
%of a planet should be proportional to its radius. Surface gravity also scales with the planet’s 
%radius, so the surface 
%pressure of a planet ($=$ surface gravity $\times$ column mass) should scale roughly as its radius squared. 
%Hence, larger planets are assumed to have thicker atmospheres whose surface pressures scale roughly as 
%$M_{p}^{2/3}$.

%       Our assumption is that larger planets can outgas non-condensibles more than smaller planets during the late stages of
%planetary formation, and therefore,
%a larger mass planet has higher N$_{2}$ partial pressure compared to a lower mass planet.  
%With this assumption, the atmospheric mass scales with planetary mass, and the column density is 
%atmospheric mass divided by the surface area of the planet. Hence, 
%larger planets are assumed to have thicker atmospheres.}
%the atmospheric column mass is proportional to the mass and inversely proportional to the square
%of the radius of the planet 
%        This ensures that the column mass of N$_{2}$ is kept constant as a function of planetary mass, and
%	Although this may be unrealistic because proportionately
%	more nitrogen is put on the smaller planet than the larger(more massive) one, 
%	a direct comparison with \cite{Kopp2013} results
%	can be made who assumed an Earth mass planet  to derive the HZs.
	Recent studies on mass-radius relationship of exoplanets have shown that mass is not directly
proportional to radius cubed; instead, it  has a more complicated relationship 
\citep{Fortney2007, Seager2010}. 
 Therefore, for our study, we used the mass and radius values of known exoplanets from the exoplanets.org 
database \citep{Wright2011} and obtained the following M-R relation:

\begin{eqnarray}
\nonumber
\frac{M_{p}}{M_{o}} &=& 0.968 \biggl(\frac{R_{p}}{R_{o}}\biggr)^{3.2}, M_{p} < 5 M_{\oplus}
\end{eqnarray}
%the surface gravities (and hence the scaling of N$_{2}$) were 
%relations  given in \cite[Page 383, Equation (12) and table 5]{Seager2010}. 
%Note that these relations match with the values
%given in Table 1 of \cite{Fortney2007} for Earth-like composition, which is our assumption for habitability.
Using this relation, the 
surface pressure in Eq.(\ref{press2}) can be written as: 
\begin{eqnarray}
\frac{P_{s}}{P_{s}^{o}} &=&  0.937 \biggl(\frac{R_{p}}{R_o} \biggr)^{2.40},   M_{p} < 5 M_{\oplus} 
\label{press3}
\end{eqnarray}
The above equation suggests that larger planets  should have thicker atmospheres. An upper limit of
$5$ M$_{\oplus}$ is motivated by the observation that planets more massive than this limit seem to 
have a steeper
slope in the M-R relation than the one predicted by \cite{Seager2010} or \cite{Fortney2007} 
for Earth-like composition.
%(see Fig.\ref{m-r}). 
%These objects have larger radii and lower densities, and do not fall along
%the terrestrial planet model curves.
%A more detailed analysis of this limit will be developed elsewhere, but 
For now we assume  that planets with masses $> 5$ M$_{\oplus}$ are not rocky.
%, and hence are beyond the scope of this study.

%	As in \cite{Kopp2013}, 
$\h2o$ and $\co2$ clouds were neglected in the model, 
but the effect of the former is accounted for by increasing the surface albedo, as done in previous 
	climate simulations by the Kasting research group
	\citep{Jacob2008, Ramirez2013}.

%	\begin{figure}[!hbp|t]
%	%%%\centering
%%	\includegraphics[width=.95\textwidth]{figures/HZ_flux_planetMass_update_1.eps}
%	\includegraphics[width=.95\textwidth]{mass-radius_ravipaper.eps}
%
%	\caption{Mass-radius space for exoplanets (blue circle) and Solar system objects
%                (red triangle). Also plotted are the model fits for terrestrial planets
%                from \cite{Seager2010} (solid green) and empirical M-R relation from \cite{WM2013} (dashed magenta).
%                The square box represents the upper mass limit ($5$ M$_{\oplus}$) for a terrestrial 
%                planet {\it assumed} in this paper.}
%	\label{m-r}
%	\end{figure}

	\section{Results}
	\label{results}

\begin{figure}[!hbp|t]
         \subfigure[]{
         \label{pn2vary}
         \includegraphics[width=.54\textwidth]{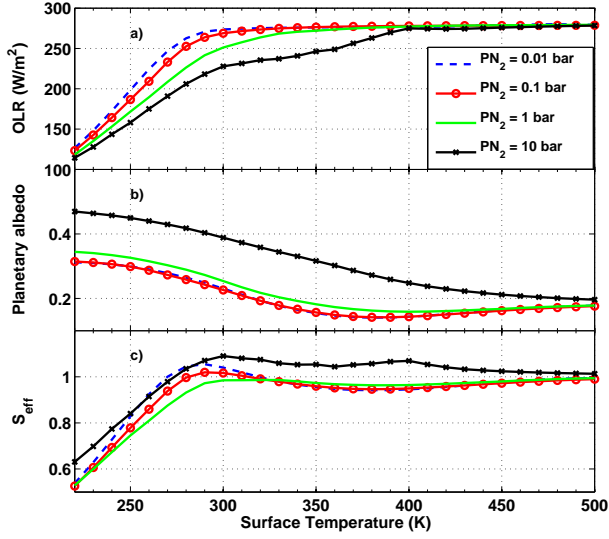}
         }
         \subfigure[]{
         \label{mevary}
         \includegraphics[width=.54\textwidth]{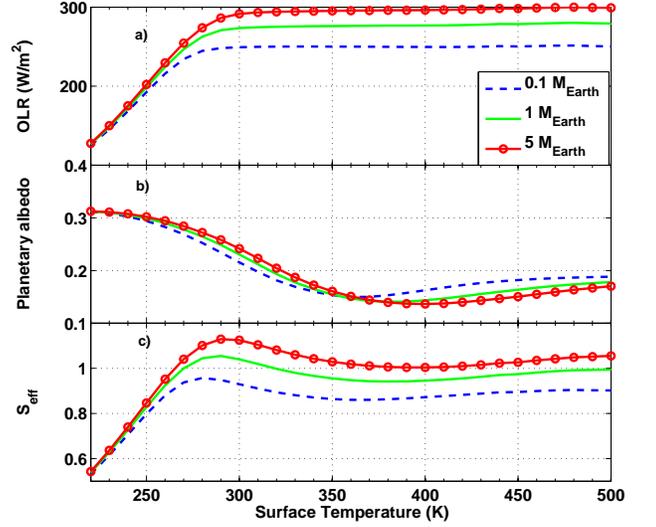}
         }
         \subfigure[]{
         \label{pn2varyohz}
         \includegraphics[width=.54\textwidth]{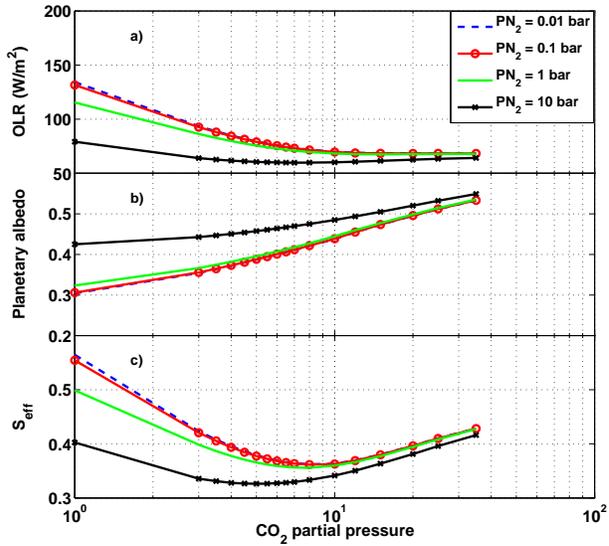}
         }
         \subfigure[]{
         \label{mevaryohz}
         \includegraphics[width=.54\textwidth]{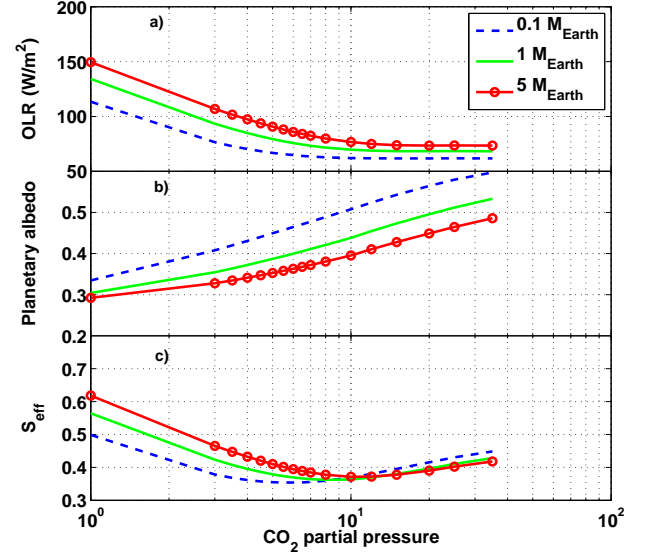}
         }

\caption{Variation in radiative fluxes and planetary albedo as a function of background N$_{2}$ partial pressure
(panels {\it (a)} and {\it (c)}) and planetary mass (panels {\it (b)} \& {\it (d)}). The top row is for the inner
edge and the bottom row is for the outer edge of the HZ. }
	\label{pn2me}
	\end{figure}

         In Fig. \ref{pn2me}, we show the variation in the calculated outgoing longwave 
radiation (OLR), planetary albedo 
and the effective solar flux ($S_{eff}$) incident on the planet as a function of the surface temperature 
(top row), and $\co2$ partial pressure (bottom row). Panels {\it a-b} and {\it c-d} correspond 
to the inner and outer edge of the HZ,
respectively. All the calculations assume a Sun-like star.
Fig. \ref{pn2vary} shows 
the case where the background N$_{2}$ partial pressure ($pN_{2}$) is varied
%Fig. \ref{pn2vary} shows the variation in these three parameters 
from $0.01 - 10$ bar for a $1$ M$_{\oplus}$ planet. At lower surface temperatures ($< 350$ K), where 
the $\h2o$ vapor is not a major constituent of the atmosphere, the net OLR is higher for lower
$pN_{2}$. The reason is that the pressure broadening by N$_{2}$ is not effective at
lower pressures, and hence results in less IR absorption and an increase in OLR. Another way to look
at it is that, to radiate the same amount of OLR, the surface temperature needs to be higher for larger
 $pN_{2}$. At higher surface temperatures ($> 350$ K), water vapor dominates the 
atmosphere, the atmosphere becomes opaque IR radiation, and the OLR asymptotes to a limiting value of 
$\sim 280$ Wm$^{-2}$.  A similar calculation performed by \cite{RayP2010} shows a distinctive peak in
the OLR for low $pN_{2}$, whereas our model does not show this feature. A possible reason could be that we
are using \cite{Ingersoll1969} formulation to calculate the adiabat, and perform a finer sublevel integration
to calculate the cold-trap accurately. Although, this feature does not affect our conclusions, a more
thorough investigation is needed to resolve these discrepancies.

The planetary albedo (second panel)
is higher for larger N$_{2}$ pressures due to the Rayleigh scattering arising from the higher amount of
non-condensable gas. The net effect of both the OLR ($F_{IR}$) and planetary albedo (or the 
net absorbed solar flux, $F_{SOL}$)
 can be combined to obtain 
$S_{eff} = F_{IR}/F_{SOL}$, shown
in the bottom panel.  The inner edge of the HZ in our model is determined by the ``runaway greenhouse limit'', 
where
the limiting OLR (or $S_{eff}$) is reached and the ocean vaporizes completely. This replaces the 
``moist-greenhouse limit'' where the stratosphere becomes wet, which defined the HZ inner edge in 
\cite{Kasting1993} and \cite{Kopp2013}. The reason is two fold: (1) Both these limits occur in our model
at $S_{eff}$ values within $2 \%$ of each other, so the difference is minimal. And (2) \cite{Leconte2013} 
predict a 
much lower tropopause temperatures than that predicted by our 1D model, due to non-grey radiative effects
and unsaturated regions that flatten the thermal profile in the troposphere. 
Consequently, their tropopause temperature can be as low as $115$ K, as compared to the $200$ K assumed 
in our inverse 1D calculations.
Further independent analysis is needed to test the robustness of this result, as non-LTE (Local Thermodynamic
Equilibrium) effects might also be important.

Since the asymptotic OLR is similar for 
different amounts of $pN_{2}$, we conclude that the inner edge of the HZ
is depends weakly on the background N$_{2}$ present in the atmosphere for a given planet mass. 
%Note that, if we consider the ``moist-greenhouse'' limit as the inner edge limit (as was the case in \cite{Kopp2013}),
%then it occurs at lower temperatures for lower N$_{2}$ pressures, due to the lack of IR absorption due to 
%pressure broadening due to N$_{2}$}

 Fig. \ref{mevary} illustrates the effect of planet mass (or gravity) on OLR, albedo and $S_{eff}$. 
Planetary masses of $0.1, 1$ and $5$ M$_{\oplus}$ are chosen to encompass the terrestrial planet range. The background
N$_{2}$ pressure is fixed at a low value of $0.01$ bar to study the effect of gravity alone with minimal contribution 
from the non-condensable gas. Fig. \ref{mevary} shows that the limiting OLR is higher for massive planets. 
This is because the $\h2o$ 
column depth is larger for the $0.1$ M$_{\oplus}$ planet owing to its low gravity, 
which increases the greenhouse effect and reduces the OLR.
The planetary albedo does not vary significantly, as the amount of N$_{2}$ present in the
atmosphere is low\footnote{For larger N$_{2}$ pressures, the albedo is higher for low mass planets  because
proportionately more nitrogen is put on the smaller planet which increases the Rayleigh scattering. But as the 
temperature increases, the albedo for all the planets asymptote to nearly the same value.}. The
net effect is that, for massive planets, $S_{eff}$ is larger compared to low mass planets. Therefore, the inner edge
of the HZ moves closer to the star for more massive planets.

%\cite{RayP2010} discussed the effect of non-condensable background gas that is
%transparent in the infrared (N$_{2}$) on the outgoing
%long-wave radiation (OLR) (see section $4.6$ in \cite{RayP2010}). Assuming a planetary surface gravity of
%$20$ ms$^{-2}$ and varying the amount of N$_{2}$, he found that the peak OLR is higher for a
%larger amount of N$_{2}$. Furthermore, this peak
%occurs at a higher temperature as the background N$_{2}$ amount is increased. This essentially
%means that for a planet with a higher amount of
%non-condensable background gas (as assumed for a $5$ M$_{E}$ planet in our model),
%one has to go to higher temperatures before water vapor completely dominates the OLR.  

Figs. \ref{pn2varyohz} and \ref{mevaryohz} show the results for the outer edge of the HZ, with the same variation in
N$_{2}$ pressures (for $1$ M$_{\oplus}$) and planetary mass (with $pN_{2}=0.01$ bar) as in Figs. \ref{pn2vary} 
and \ref{mevary}. Fixing the surface temperature
at $273$ K, we varied the $\co2$ partial pressure from $1$ to $35$ bars and calculated the corresponding 
radiative fluxes and planetary albedos. As with the inner edge case, less absorption occurs at low N$_{2}$
 pressures because of ineffective pressure broadening, and this
 results in an increase in the OLR. 
This effect is augmented by an increase in planetary albedo at high $pN_{2}$, resulting in decreased absorption of 
solar radiation. Thus, towards the left-hand side of Fig. \ref{pn2varyohz}, where $pCO_{2}$ is low, 
$S_{eff}$ is considerably higher at low $pN_{2}$.
%This has little effect on the HZ outer edge for low $pN_{2}$, 
%however, because 
The HZ outer edge (the 'maximum greenhouse' limit) is determined by the minimum in $S_{eff}$. 
This boundary occurs at lower $S_{eff}$ (i.e, further from the star) for large $pN_{2}$ ($10$ bar).
For low enough $pN_{2}$ values, that minimum is governed by $\co2$, not $N_{2}$ \citep{vonParis2013}.
Hence, the outer edge of the HZ does not change significantly for these low $N_{2}$ pressures.

\begin{figure}[!hbp|t]
         \subfigure[]{
         \label{pn2scaleihz}
         \includegraphics[width=.50\textwidth]{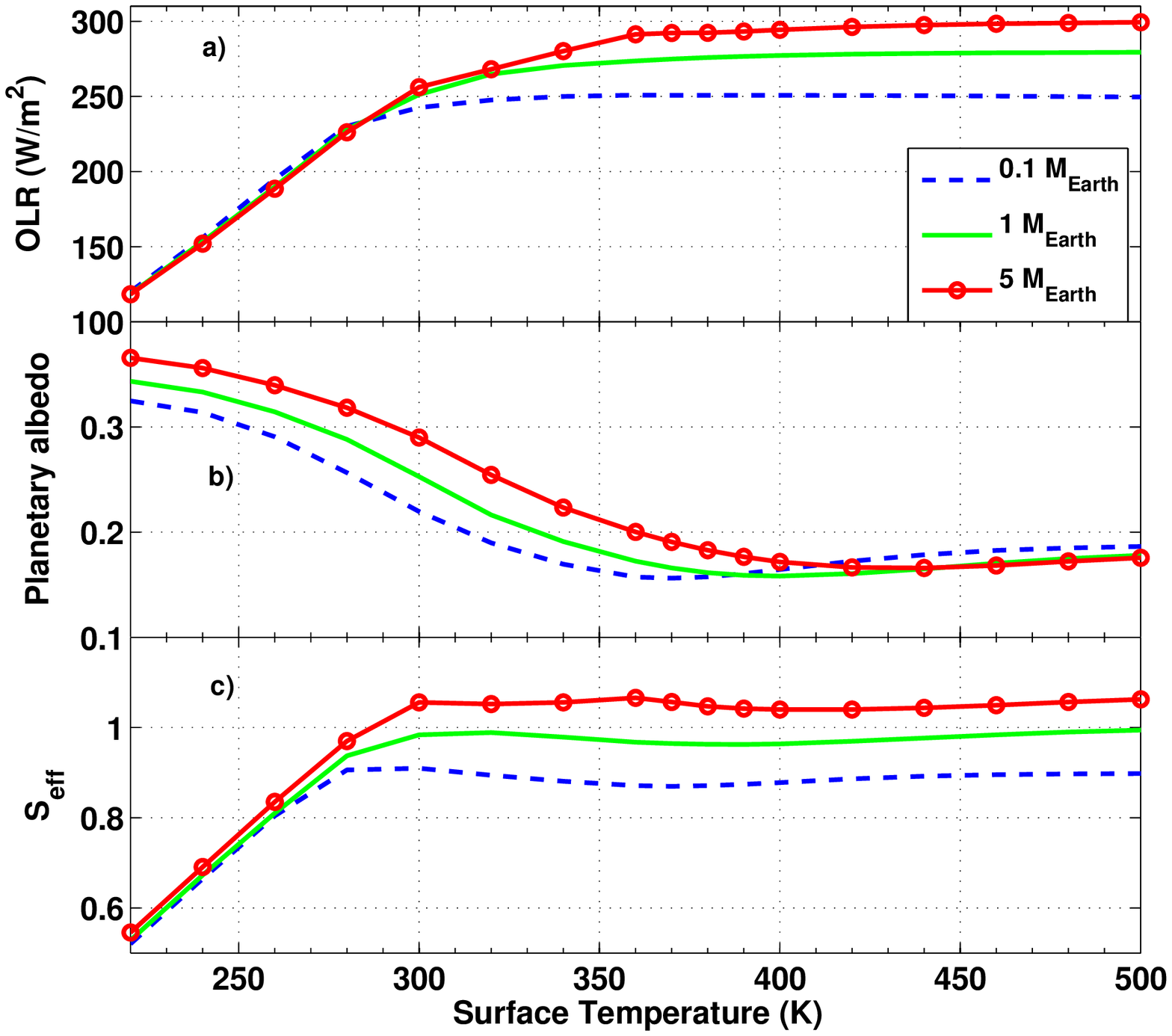}
         }
         \subfigure[]{
         \label{pn2scaleohz}
         \includegraphics[width=.50\textwidth]{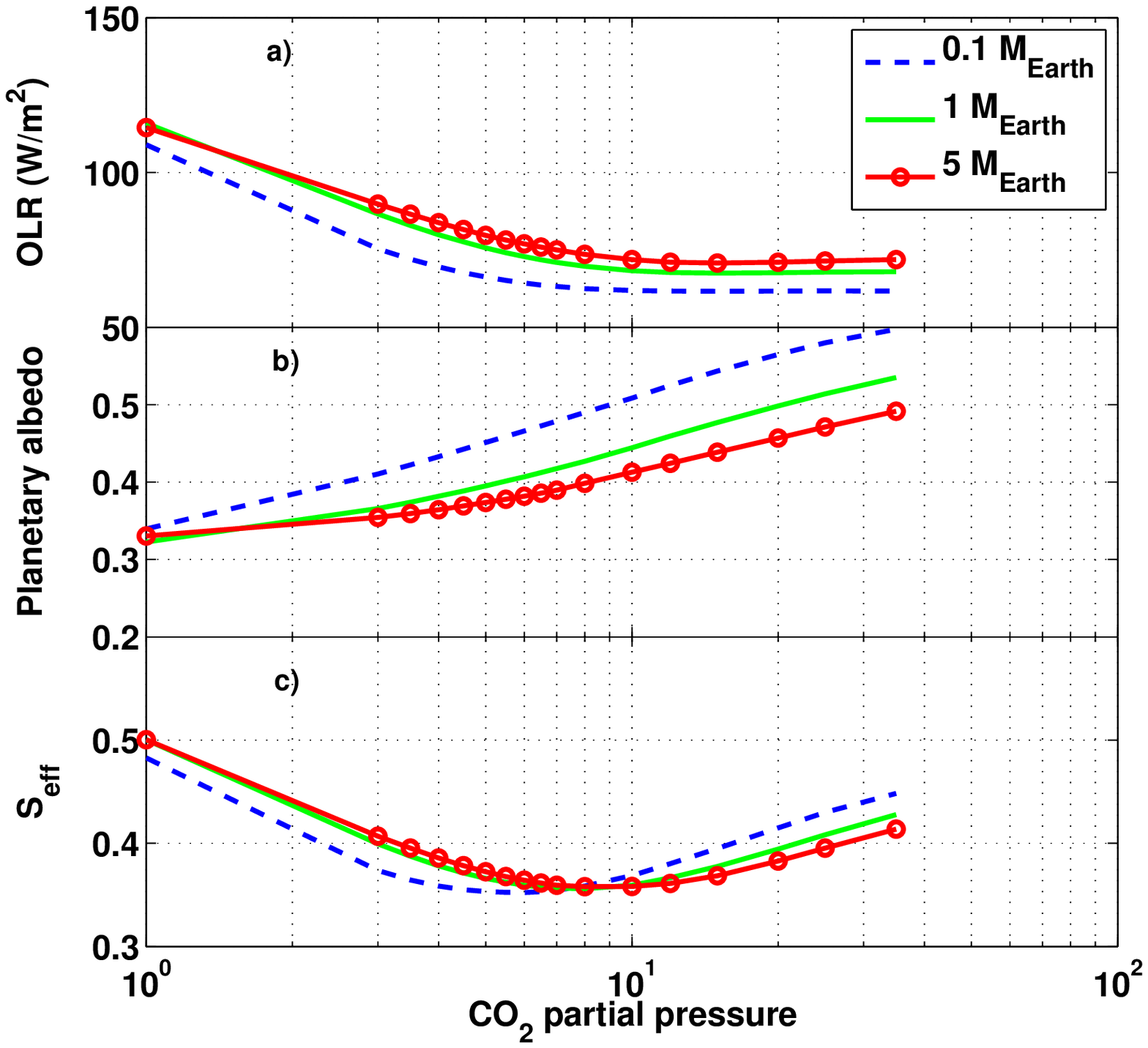}
         }

	\caption{ Similar to Fig. \ref{pn2me}, but here $pN_{2}$ is scaled according to the planet mass 
as described by Eq.(\ref{press3}). The net effect is that the inner
edge of the HZ (left-bottom panel) moves closer to the star for a massive planet and the outer edge of the HZ (right-bottom panel) changes 
little.}
	\label{fluxpMass}
	\end{figure}

 As mentioned in section \ref{sec2}, we considered a third case where the background N$_{2}$ pressure is scaled 
according to the planetary mass. We consider this case to be a more realistic estimate for the non-condensable 
background gas concentration in a planetary atmosphere for the reasons 
outlined in section \ref{sec2}. Fig. \ref{fluxpMass} shows the inner (left panel) and outer 
(right panel) edge calculations for this case 3.
These results can be understood by recognizing that they represent a combination of various cases shown in Fig. \ref{pn2me}.
For example, Fig. \ref{pn2vary} shows that increasing $pN_{2}$ for a given planet mass shifts the peak OLR to higher temperatures
due to pressure broadening (compare the $2$ bar case with $0.01$ bar).  Also, Fig. \ref{mevary} illustrates that the OLR is 
larger for a more massive planet due to smaller atmospheric column depth (for a given $pN_{2}$), and hence results in less IR absorption.
Both these effects can be seen in Fig. \ref{pn2scaleihz}, where both the planet mass  and $pN_{2}$ are varied: The peak OLR shifts
to higher temperatures because $pN_{2}$ is scaled, and the $5$ M$_{\oplus}$ planet has a higher OLR than a 
$0.1$ M$_{\oplus}$ planet which is
a direct consequence of the results shown in Fig. \ref{mevary}.

Similar reasoning can be applied to the outer edge of the HZ (Fig. \ref{pn2scaleohz}). We showed in Figs. \ref{pn2varyohz} and 
\ref{mevaryohz} that,
there is not a signficant change in $S_{eff}$ for different planetary masses due to the competing effects of the greenhouse effect
of $\co2$ and the planetary albedo. This is reflected in the bottom panel of Fig. \ref{pn2scaleohz}. 
Since the inner edge moves closer to the star
        for the super-Earth planet, while the outer edge changed little, we can conclude that
         larger (more massive) planets have  wider habitable zones than do small
        ones.

We should note that we found an error in our previously derived $\h2o$ IR coefficients,
which caused us to  underestimate ($\sim 4 \%$) the strength of the absorption by these gases at the inner edge.
 We have now 
corrected this error.
As a result, the runaway greenhouse 
limit moves to lower stellar fluxes, and Earth now falls right on this limit 
suggesting that Earth should be in the runaway greenhouse state. This reflects our 1D model's inability
to realistically account for variations in relative humidity and clouds, 
which move IHZ to higher stellar fluxes, as discussed earlier.

\subsection{Variation of HZs With Planetary Mass}
\label{discuss}

 The results from the previous section can be extended to stars with different $T_{eff}$. Specifically, we use the results
from $pN_{2}$ scaling with planetary mass to derive various HZ limits for stars with $2600$ K $ <= T_{eff} <= 7200$ K.

\begin{figure}[!hbp|t]
	\includegraphics[width=.90\textwidth]{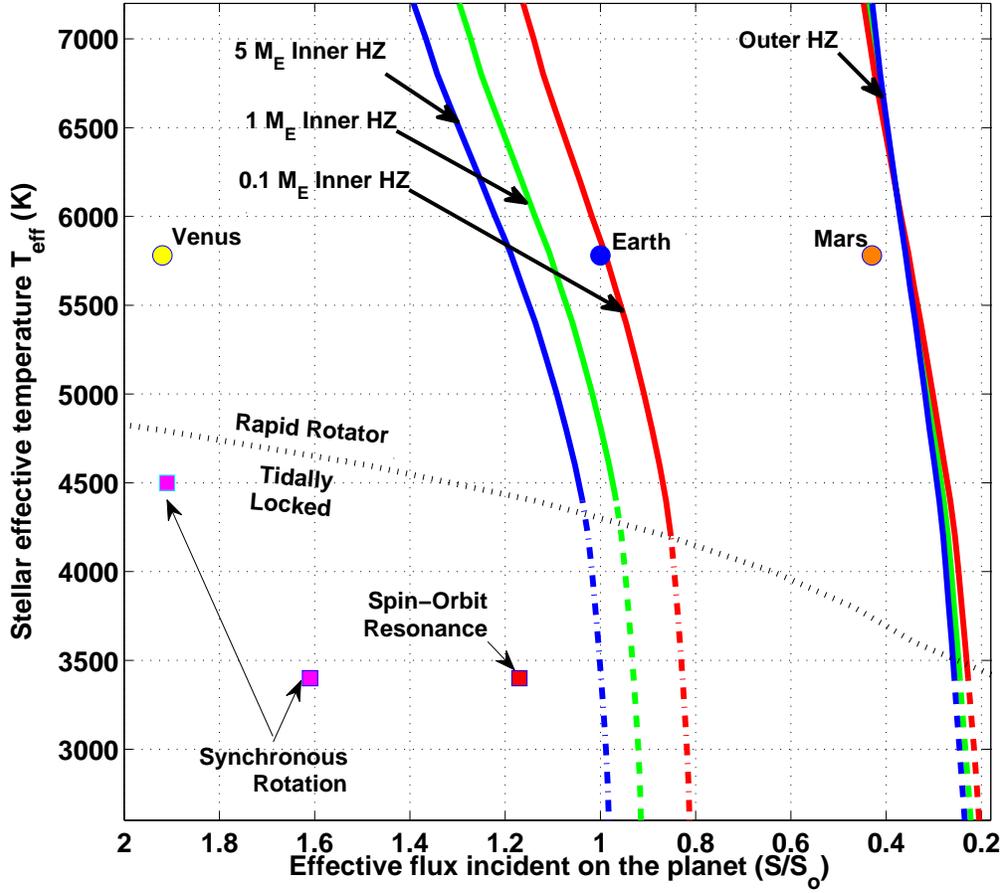}
	\caption{HZ limits for different planetary masses. The inner HZ for Earth (green curve) is scaled with the Leconte et al.(2013) inner
HZ for the Sun, using our runaway greenhouse limit. The outer HZ  is the maximum greenhouse limit. 
The tidal locking radius (black dot), assuming a $4.5$ Gyr tidal locking timescale, separates rapidly rotating planets with tidally
locked ones.
The inner-edge limits for tidal-locked planets around cool stars from \cite{Yang2013} are also shown (colored squares). }
%Further 3D model studies are needed to verify these results.}
%Planets that are orbiting cool stars 
%may have wider HZs \citep{Yang2013} if they are synchronously rotating (red solid). Also
%shown are Solar system's terrestrial planets (colored circles). } 
	\label{3dhz}
	\end{figure}

%  As discussed in the introduction, for both tidally locked and rapidly rotating planets, 
%3D GCM studies suggest that the inner edge of the HZ is closer to the star than predicted by 1D models.
By integrating the 1D and 3D model results, we have constructed the various HZ limits in Fig. \ref{3dhz}. 
For rapidly rotating planets like the Earth, we scale the \cite{Leconte2013} inner edge limit with our value of the 
runaway greenhouse limit for different stars, and obtain a 
``conservative'' estimate of the inner edge of the HZ (green curve). Note that Earth is well inside the HZ in this figure, 
as it should be, because the \cite{Leconte2013} runaway greenhouse limit occurs at a higher stellar flux. 

  For cool stars ($T_{eff} \le 4500$ K), the inner HZ is a function of tidal locking radius 
(\cite{Edson2011}, dashed and solid black line in Fig. \ref{3dhz} assuming $4.5$ Gyr tidal lock timescale). 
The \cite{Yang2013} GCM models considered an M-star with
$T_{eff} = 3400$K and a K-star with $T_{eff} = 4500$K. We show their model results in
%extrapolated their synchronously 
%rotating planet limits to stars with $T_{eff} <3400$K  
Fig.\ref{3dhz} for both synchronously rotating and a 6:1 spin-orbit resonance case. This result needs to be verified
with further studies.
%Yang et al. (2013) also performed a climate calculation for a planet in a 6:1 spin-orbit resonance around a 3400 K M star. The inner edge for this type of planet is well outside the synchronous limit (magenta square).

%Following \cite{Kasting1993, Kopp2013}, 
A conservative estimate of the outer edge of the HZ is defined by the maximum 
greenhouse limit (blue solid curve). 
%which is the greatest distance at which a planet can be warmed above freezing by a $\co2$-$\h2o$ atmosphere.
 The actual outer edge could be further out if additional greenhouse gases (e.g., H$_{2}$) are present \citep{PG2011}.
% As discussed in previous section, 
The outer edge does not vary
significantly for different planetary masses.
 The change in the stellar flux at the inner edge of the HZ, compared to Earth,
  is $\sim 10 \%$ for $0.1$ M$_{\oplus}$ and $\sim 7 \%$ for $5$ M$_{\oplus}$ in our model.

%For ease of calculation, 
We provide parametric equations to calculate HZs:
%of both
%the conservative and optimistic HZ fluxes  below:
%in the range $2600$ K $ \le T_{eff} \le 7200$ K:
\begin{eqnarray}
\label{hzeq}
S_{eff} &=&  S_{eff\odot} + aT_{\star} + bT_{\star}^2 + cT_{\star}^3
                + dT_{\star}^4 
\end{eqnarray}
where $T_{\star} = T_{eff} - 5780$ K and the coefficients are listed in Table \ref{table1}. The corresponding
habitable zone distances can be calculated using the relation:
\begin{eqnarray}
d &=& \biggl(\frac{L/L_{\odot}}{S_{eff}}\biggr)^{0.5} \mathrm{AU}
\label{dhz}
\end{eqnarray}
where $L/L_{\odot}$ is the luminosity of the star compared to the Sun.
        
	\section{Conclusions}
	\label{conclusions}
	The HZ boundaries change
	as a of function planetary mass and the amount of background N$_{2}$ gas. 
	The conservative HZ limits for  more massive planets should be wider 
than those for low mass planets if the atmospheric column depth scales with planet radius, as assumed here.
 The results summarized here are  only a step towards 
a more comprehensive analysis of HZ boundaries. 
%There may be an abrupt change in the width of 
%the HZ from rapidly rotating to tidally locked planets. This part of the parameter space needs to be
%explored more throroughly.
Further work with 3D climate models will be needed to  accurately calculate the habitable zones around different types of stars.
%	In order to facilitate the calculation of HZs presented in this paper, we have derived  parametric
%expressions for HZ fluxes as a function of planetary masses and stellar effective temperatures.
%	A FORTRAN  code that calculates the HZ limits is provided in the online version of the journal.

	A FORTRAN code is available with the online version of the paper. An interactive webpage to obtain HZs is available at:
        \url{http://www3.geosc.psu.edu/~ruk15/planets/} or at 
	\url {http://depts.washington.edu/naivpl/content/hz-calculator}. 

	\acknowledgements

%	The authors acknowledge the Research Computing and Cyberinfrastructure
%	unit of Information Technology Services at The Pennsylvania State
%	University for providing advanced computing resources and services that
%	have contributed to the research results reported in this paper. {\url {
%	http://rcc.its.psu.edu}}.  This work was also facilitated through the use of 
%	advanced computational, storage, and networking infrastructure provided by the 
%	Hyak supercomputer system, supported in part by the University of Washington eScience Institute.
%	This research has made use of the Exoplanet Orbit Database
%	and the Exoplanet Data Explorer at exoplanets.org.

        The authors thank an anonymous reviewer whose comments greatly improved the manuscript.
	R. K, R. R, J.F.K and S.D.G gratefully acknowledge funding from NASA Astrobiology
	 Institute's  Virtual 
	Planetary Laboratory lead team, supported by NASA under cooperative agreement
	NNH05ZDA001C, and the Penn State Astrobiology Research Center.
	V.E. acknowledges the support of the European Research Council (Starting Grant 209622: E3ARTHs).
The Center for Exoplanets and Habitable Worlds is supported by the
	Pennsylvania State University, the Eberly College of Science, and the
	Pennsylvania Space Grant Consortium. R.K and R.R contributed equally to this work.

\begin{center}
\begin{threeparttable}[h!]
\caption{Coefficients to be used in Eq.(\ref{hzeq}). The coefficients for recent Venus, Maximum
Greenhouse and early Mars are same for all the planetary masses. For $5$ M$_{\oplus}$ and 
$0.1$ M$_{\oplus}$, the background N$_{2}$ pressure is scaled according to the planetary mass.
% to calculate habitable stellar fluxes,
%for stars with
%$2600 \le T_{eff} \le 7200$ K. 
An ASCII file containing these coefficients can be downloaded
 in the electronic version of the paper.}
\vspace{0.1 in}
\centering
\begin{tabular}{|c|c|c|c|c|}
\hline
%\multicolumn{4}{|c|}{~~~~~~~~F0 ~~~~~~G2~~~~~K2~~~~~M0~~~~~M8} \\
%\hline
%\cline{2-3}
%\cline{4-5}
%\centering
Constant& Recent & Runaway & Maximum & Early\\
&Venus&Greenhouse&Greenhouse & Mars\\
\hline
$S_{eff\odot}$ ($1$ M$_{\oplus}$) & 1.776& 1.107 &0.356& 0.32 \\
&&&&\\
$S_{eff\odot}$ ($5$ M$_{\oplus}$) & -- & 1.188 & --& -- \\
&&&&\\
$S_{eff\odot}$ ($0.1$ M$_{\oplus}$) & --& 0.99 &--& -- \\
\hline
&&&&\\
$a$ ($1$ M$_{\oplus}$) & $2.136 \times 10^{-4}$ & $1.332 \times 10^{-4}$  &$6.171 \times 10^{-5}$ & $5.547 \times 10^{-5}$  \\
&&&&\\
$a$ ($5$ M$_{\oplus}$) & -- & $1.433 \times 10^{-4}$  &-- & --  \\
&&&&\\
$a$ ($0.1$ M$_{\oplus}$) & -- & $1.209 \times 10^{-4}$  &-- & --  \\
&&&&\\
\hline
$b$ ($1$ M$_{\oplus}$)& $2.533 \times 10^{-8}$& $1.58 \times 10^{-8}$ &$1.698 \times 10^{-9}$ & $1.526 \times 10^{-9}$\\
&&&&\\
$b$ ($5$ M$_{\oplus}$)& --& $1.707 \times 10^{-8}$ &-- & --\\
&&&&\\
$b$ ($0.1$ M$_{\oplus}$)& --& $1.404 \times 10^{-8}$ &-- & --\\
&&&&\\
\hline
$c$ ($1$ M$_{\oplus}$)& $-1.332 \times 10^{-11}$ & $-8.308 \times 10^{-12}$ & $-3.198 \times 10^{-12}$ & $-2.874 \times 10^{-12}$\\
&&&&\\
$c$ ($5$ M$_{\oplus}$)& -- & $-8.968 \times 10^{-12}$ & -- & --\\
&&&&\\
$c$ ($0.1$ M$_{\oplus}$)& -- & $-7.418 \times 10^{-12}$ & -- & --\\
&&&&\\
\hline
$d$ ($1$ M$_{\oplus}$)& $-3.097 \times 10^{-15}$ &$-1.931 \times 10^{-15}$& $-5.575 \times 10^{-16}$ & $-5.011 \times 10^{-16}$\\
&&&&\\
$d$ ($5$ M$_{\oplus}$)&-- &$-2.084 \times 10^{-15}$& -- & --\\
&&&&\\
$d$ ($0.1$ M$_{\oplus}$)& --&$-1.713 \times 10^{-15}$& -- & --\\
&&&&\\
\hline
\end{tabular}
\label{table1}
\end{threeparttable}
\end{center}

\end{document}